\documentclass[10pt]{article}
\usepackage[colorlinks=true,linkcolor=blue,breaklinks=True,citecolor=brown,urlcolor=blue]{hyperref}
\usepackage{booktabs}
\usepackage{censor}
\usepackage{booktabs}
\usepackage{tabularx}
\usepackage{amsmath}
\usepackage{cleveref}
\usepackage{todonotes}
\usepackage{multirow}
\crefformat{section}{\S#2#1#3}
\crefformat{subsection}{\S#2#1#3}
\crefformat{subsubsection}{\S#2#1#3}
\usepackage{caption}

\usepackage[table]{xcolor}
\usepackage{colortbl}
\definecolor{palegreen}{RGB}{221,235,221}
\usepackage{enumitem}
\setlist[itemize]{noitemsep, topsep=0pt}

\usepackage[font={small,sf,bf},labelsep=period]{caption}

\usepackage[letterpaper]{geometry}
\usepackage{hicss51}
\usepackage{times}
\usepackage[none]{hyphenat}
\usepackage{url}
\usepackage{latexsym}
\usepackage{minted}
\usepackage{indentfirst}
\usepackage{graphicx}
\graphicspath{{images/}}

\title{LLMs in Cybersecurity: Friend or Foe in the Human Decision Loop?}

\author{
  Irdin Pekaric \\
  Department of Computer Science\\and Information Systems\\University of Liechtenstein\\
  {\underline{irdin.pekaric@uni.li}} \\\And
  Philipp Zech\\
  Department of Computer Science\\University of Innsbruck\\
  {\underline{philipp.zech@uibk.ac.at}} \\\And
  Tom Mattson \\
  Robins School of Business\\University of Richmond\\
  {\underline{tmattson@richmond.edu}} \\
}

\usepackage[
    style=apa,
  ]{biblatex}
\date{}

\begin{document}
\maketitle
\begin{abstract}
Large Language Models (LLMs) are transforming human decision-making by acting as cognitive collaborators. Yet, this promise comes with a paradox: while LLMs can improve accuracy, they may also erode independent reasoning, promote over-reliance and homogenize decisions. In this paper, we investigate how LLMs shape human judgment in 
security-critical contexts. Through two exploratory focus groups (unaided and LLM-supported), we assess decision accuracy, behavioral resilience and reliance dynamics. Our findings reveal that while LLMs enhance accuracy and consistency in routine decisions, they can inadvertently reduce cognitive diversity and improve automation bias, which is especially the case among users with lower resilience. In contrast, high-resilience individuals leverage LLMs more effectively, suggesting that cognitive traits mediate AI benefit.

\end{abstract}

\subsubsection*{Keywords: }

Large Language Models, Cybersecurity, Human Decision-Making, Human-AI Collaboration, Decision Reliability

\section{Introduction}
\label{sec:introduction}

Large Language Models (LLMs) have rapidly emerged as integral tools in domains  
characterized by complex and high-stakes decision-making, including cybersecurity, healthcare, law and finance. These models enhance human analytical  
capabilities by facilitating the parsing of complex information, automating  
knowledge extraction and synthesis, and delivering context-sensitive recommendations (\cite{zhang2025llms, ding2024large}). Beyond reducing cognitive load, LLMs  
contribute to increased decision accuracy and speed by generating structured,  
domain-specific responses in real time (\cite{lu2024information,fuegener2021borgs}). 
In cybersecurity, they are currently employed in functions such as  
threat intelligence aggregation, vulnerability assessment and phishing  
detection (\cite{cohen2025human,gallagher2024phishing}), and are increasingly combined with established approaches for model-based security analysis and vulnerability management (\cite{groner2023model,pekaric2021vulnerlizer,witte2022towards}). Similarly, in  
healthcare, LLMs assist clinicians through AI-generated treatment suggestions  
that support diagnostic and therapeutic decisions (\cite{jacobs2021antidepressants,jussupow2021diagnosis}). 
These trends reflect a broader  
transformation in human–machine collaboration: rather than replacing human  
judgment, LLMs increasingly operate as cognitive partners. This form of  
collaboration, leverages the computational  
scale and precision of machines together with human contextual insights, ethical  
discernment and domain expertise (\cite{shen2025irrationality,decremer2021kasparov}). 
When effectively integrated, such systems have the potential to  
surpass the capabilities of both human-only and AI-only decision processes  
(\cite{lu2024information,fuegener2021borgs}).

While LLMs demonstrably improve decision speed and accuracy in isolated tasks,  
emerging empirical and theoretical evidence highlights several cognitive and  
systemic drawbacks when individuals consistently rely on algorithmic guidance.  
Notably, repeated exposure to confident LLM-generated suggestions may erode  
individual decision autonomy by diminishing what has been termed ``unique human  knowledge'' --- the distinctive perspectives and heuristics that individuals bring  
to judgment processes (\cite{fuegener2021borgs,lu2024information}). Over time,  
human decisions begin to conform more closely to model outputs, reducing the  
cognitive diversity essential to effective problem solving and organizational  
resilience. This convergence effect undermines the foundational principle of  
complementarity in human-AI collaboration, wherein humans and machines  
contribute distinct and mutually reinforce capabilities (\cite{fuegener2021borgs}). 
In collective settings, such as group decision-making or ``wisdom of  
crowds'' scenarios, this effect is particularly detrimental. Studies show that  
uniform access to AI advice can suppress variability in reasoning across  
individuals, thereby reducing the accuracy and robustness of aggregated  
decisions (\cite{page2007diversity,hong2004groups}). Moreover, LLMs can induce  
automation bias, wherein users overly defer to AI outputs --- even in cases of low  
confidence or erroneous recommendations (\cite{lu2024information}). Without  
appropriate design safeguards, reliance on LLMs may thus degrade decision  
quality, particularly under uncertainty or adversarial conditions (\cite{shen2025irrationality, schroer2025dark}).

These concerns give rise to a central and timely research question:
\begin{quote}
How can  
human–LLM interaction be designed to preserve the cognitive strengths and  
judgment independence of human decision-makers while effectively leveraging the  
computational capabilities of large language models?
\end{quote}
More specifically, what  
mechanisms can support sustained decision diversity, behavioral resilience and  
autonomy in LLM-assisted decision-making, particularly in high-stakes contexts  
such as cybersecurity? Addressing these questions requires a deeper  
understanding not only of the conditions under which LLMs enhance performance,  
but also of the scenarios in which they impair independent reasoning,  
especially under cognitive stress or adversarial influence.

To investigate these issues, we conduct an exploratory focus group study that  
examines the behavioral implications of LLM-supported decision-making in  
cybersecurity contexts. Our experimental design includes two conditions: a  
control group making decisions without AI assistance (unaided) and a group  
receiving support from an LLM (LLM-supported). Participants complete domain-relevant security tasks of  
varying complexity, enabling comparative analysis of accuracy, autonomy, and  
reliance behaviors across conditions. In parallel, we assess individual  
resilience using the Brief Resilience Scale (BRS) (\cite{smith2008brief}) to explore how psychological  
factors mediate the human response to AI assistance. This design allows us to  
evaluate not only the performance effects of LLM integration, but also its  
impact on behavioral resilience, judgment diversity and error propagation.  
This study contributes to the literature on human–AI collaboration in three  
ways:
\begin{enumerate}
    \item \textit{Empirical insight into LLM influence on human judgment:} We 
        provide one of the first controlled experimental studies to evaluate how 
        LLMs affect decision reliability, independence, and behavioral resilience 
        in the cybersecurity domain.
    \item \textit{Behavioral resilience as a novel lens:} We employ 
        the concept of behavioral resilience to capture the degree to which human decision 
        makers maintain autonomy and resist over-reliance when engaging with LLMs.
    \item \textit{Design implications for human-in-the-loop AI systems:} Our findings 
        offer actionable recommendations for designing LLM interfaces and interaction 
        protocols that promote human diversity and reduce automation bias in hybrid 
        decision workflows.
\end{enumerate}

The remainder of this paper is organized as follows: \cref{sec:background} examines related work on the integration of LLMs in cybersecurity systems and their effect on decision reliability; \cref{sec:methodology} describes the applied method; \cref{sec:results} provides the quantitative, qualitative and BRS-related results; \cref{sec:discussion} discusses key findings and implications; \cref{sec:conclusion} concludes this paper.

\section{Related Work}
\label{sec:background}

LLMs are increasingly seen as promising tools in 
cybersecurity—not only for automating routine tasks but also for augmenting 
human decision-making in critical environments. This section reviews prior 
work across three converging themes: the integration of LLMs in cybersecurity 
systems, their effect on decision reliability and the dual role of technical 
and behavioral resilience in shaping security outcomes.

\textcite{zhang2025llms} deliver a comprehensive review of LLMs in 
cybersecurity, analyzing over 300 studies across domains such as threat 
intelligence, vulnerability detection, and secure code generation. They 
underscore critical challenges including domain-specific adaptation, adversarial 
robustness and scalability. They also chart a development roadmap toward 
production-grade, security-focused LLM architectures. 

\textcite{ding2024large} frame LLMs as enablers of cyber resilience, 
organizing their impact across five pillars: security posture, data protection, 
user awareness, network defense and automation. Through comparative case 
studies, they demonstrate how LLM-enhanced systems outperform classical models 
in both detection precision and recovery agility—particularly in layered, 
adaptive defense settings.

\textcite{khadka2025human} turn attention to the human side of 
cybersecurity, critiquing overly technical paradigms and proposing a 
socio-technical framework that centers psychological resilience, adaptive 
learning and ethical AI. Their human-centric cybersecurity framework prioritizes 
emotional intelligence, gamified awareness training and decision-support 
integration as cultural anchors for organizational resilience.

From an operational intelligence standpoint, \textcite{cohen2025human} 
present an empirical model of human–AI collaboration in threat detection. Using a 
reciprocal learning framework, human experts and LLMs iteratively and jointly classify 
hacker forum data. Their approach not only improves classification accuracy over 
time but also refines the experts’ conceptual understanding of attacker profiles.

On the engineering front, \textcite{bedoya2024enhancing} explore the 
synthesis of LLMs and Security Chaos Engineering within DevSecOps 
workflows. They propose a design-time integration strategy that combines 
attack-defense tree modeling with LLM-generated threat simulations to 
proactively expose hidden vulnerabilities before deployment.

\textcite{sarker2024explainable} address explainability in 
LLM-supported cybersecurity environments, particularly within Digital Twins. 
Their taxonomy emphasizes the trade-offs among automation, intelligence and 
trustworthiness, which they frame as ``CyberAIT''. In their work, they highlight the importance of 
transparency and human-aligned reasoning in mission-critical systems.

\textcite{nguyen2024utilizing} propose a human-in-the-loop phishing 
defense mechanism that leverages LLMs to generate personalized, actionable 
warnings. Their approach integrates user-specific insights into content analysis, 
achieving perfect classification with zero false positives and improving user 
responsiveness under threat.

Finally, \textcite{chen2024survey} conduct a broad survey of LLM use in 
cyber threat pipelines, focusing on NLP-driven detection stages. They document 
the technical limitations of fully automated systems and argue for hybrid 
human–AI detection architectures to reduce false positives and adapt to evolving 
attack vectors.

Taken together, these studies offer compelling evidence that LLMs can reinforce 
both technical and human dimensions of cybersecurity resilience. Yet they also 
highlight persistent gaps. Notably absent is controlled, empirical 
research on how LLMs influence human decision accuracy under cognitive stress. 
This study addresses that gap by experimentally comparing participants’ task 
performance --- with and without LLM support --- while also measuring individual 
resilience. The findings contribute to a more grounded understanding of AI’s 
role in human-in-the-loop cybersecurity contexts.

\section{Research Design and Methods}
\label{sec:methodology}

To investigate human-LLM interactions, we conducted two exploratory focus groups. Our focus groups allowed us to gain in-depth understanding of individuals' decision quality, perceptions, attitudes, and beliefs pertaining to the usefulness of LLMs for different information security issues. Our goal was to better understand the complexities associated with human-LLM interactions rather than generalize our findings to large populations (\cite{morgan1996focusgroups}), which made focus groups an effective choice for our paper. Focus-groups are also particularly appropriate for exploratory research without a priori hypotheses such as ours.

To facilitate our focus groups, we first developed a template to guide our sessions. The template included common security tasks performed by security professionals. To ensure domain relevance and realism, we aligned the task design with topics covered in the CompTIA Security+ certification exam\footnote{\url{https://www.comptia.org} - the exam is available both as a web and a mobile version, as indicated by~\textcite{pekaric2025mobile}}, which outlines foundational competencies in threat detection, incident response and security best practices. We included two easy, two medium, and two difficult tasks as determined by security academic researchers. The tasks were related to phishing, password management, suspicious foreign country network logins, log file anomaly detection, exploit identification, and system-level resilience strategies for common security threats. We asked our participants to perform tasks, make decisions, and discuss their rationale. We wanted to learn how the use (or lack thereof) of LLMs impacted the quality of and individuals’ perceptions of LLMs for particular security tasks. We refined and reviewed our template along with the six tasks that guided each focus group session before implementing it.

Additionally, we measured each participant's individual resilience, because individual-level resilience influences how humans perceive and respond to challenges, stress, and uncertainty, which is relevant to security professionals responding to dynamic threats and vulnerabilities. Individual resilience includes important elements such as emotional regulation, adaptability, problem solving, and optimism. All of which are important factors in human decision making. To capture this construct, we adopted a six-item individual resilience scale from~\textcite{smith2008brief}, which captures individuals’ capacity to bounce back from stress and adversity that often happen in security professionals’ day-to-day activities. Each item on the scale had five answer options including strongly disagree (1), disagree (2), neutral (3), agree (4), and strongly agree (5). The final resilience values for each participant are the sum of the six items on the Likert scale.

We recruited convenience and purposeful samples of master’s degree students who had backgrounds in cybersecurity to participate in our focus groups. The first focus group had our participants complete and discuss tasks without the help of any LLM. We had 11 participants with an average age of 31.1 ($\sigma = 9.9$) and an average individual resilience score of 18.4 ($\sigma = 2.5$). The second focus group used the same template, but allowed participants to perform the tasks with the aid of any LLM. We had 10 participants with an average age of 26.6 ($\sigma = 3.4$) and an average individual resilience score of 17.9 ($\sigma = 0.88$).  We release the templates for unaided and LLM-supported groups in our open-source repository: \url{https://github.com/irdin-pekaric/hicss2026llm}.
 
\section{Results}
\label{sec:results}

In the following we provide our results on the differences between the unaided and LLM-supported groups. To this end, we conduct quantitative analysis of closed-ended questions, qualitative analysis of open-ended questions and statistical analysis of the impact of the personal resilience on the performance for each of the two groups. 

\subsection{Quantitative Analysis}

We now report the results of the quantitative analysis for tasks that have different difficulties.

\begin{figure}[h!]
    \centering
    \includegraphics[width=\columnwidth]{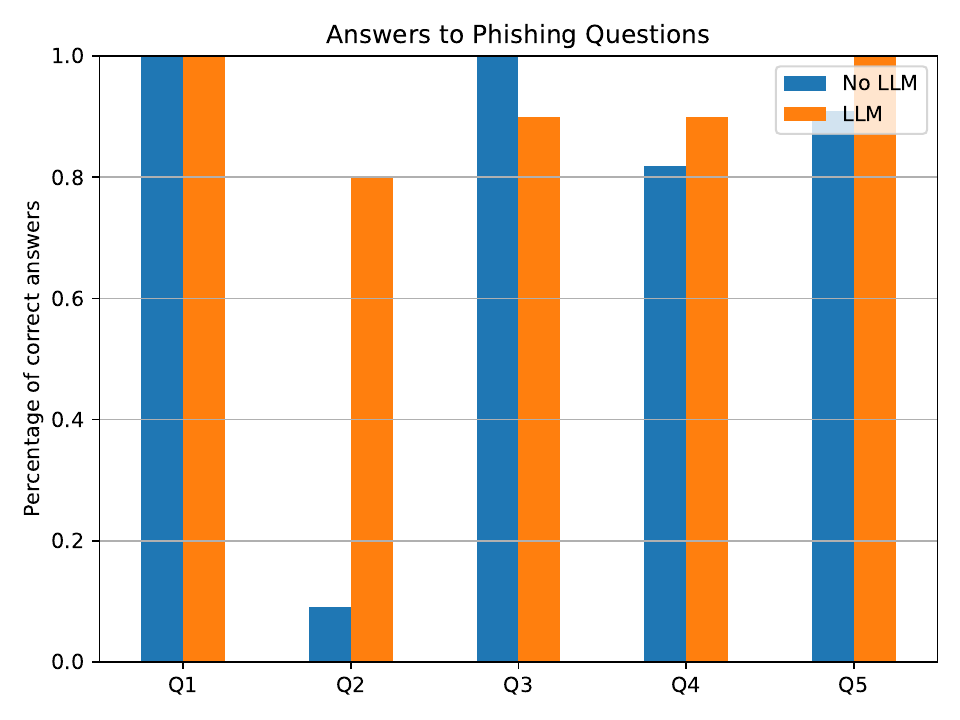}
    \caption{Proportion of correct answers by question, comparing unaided and LLM-supported groups.}
    \label{fig:phishing_comparison}
\end{figure}

\noindent \textbf{Phishing (Easy Difficulty) Questions} - The results reveal a consistent difference in participants’ ability to correctly answer phishing-related questions, depending on whether they received LLM support (see Figure \ref{fig:phishing_comparison}). Across all five questions, which were designed to assess recognition of phishing and non-phishing content, participants aided by an LLM outperformed those without support. In the non-LLM group, the proportion of correct answers ranged from 0.64 to 1.00, whereas the LLM group showed higher consistency, with accuracy ranging from 0.88 to 1.00. The most notable gap was observed in Question 2, a non-phishing item where ``No'' was the correct response: only 18\% of the non-LLM group answered correctly, compared to 88\% in the LLM group. This suggests that LLM assistance can help to improve participants’ ability to correctly identify legitimate messages and avoid false positives. Accuracy for the other phishing questions remained high in both groups, but still showed modest improvements with LLM support. Overall, these findings indicate that LLMs may enhance human performance in distinguishing phishing threats from safe communications.

\begin{figure}[H]
    \centering
    \includegraphics[width=\columnwidth]{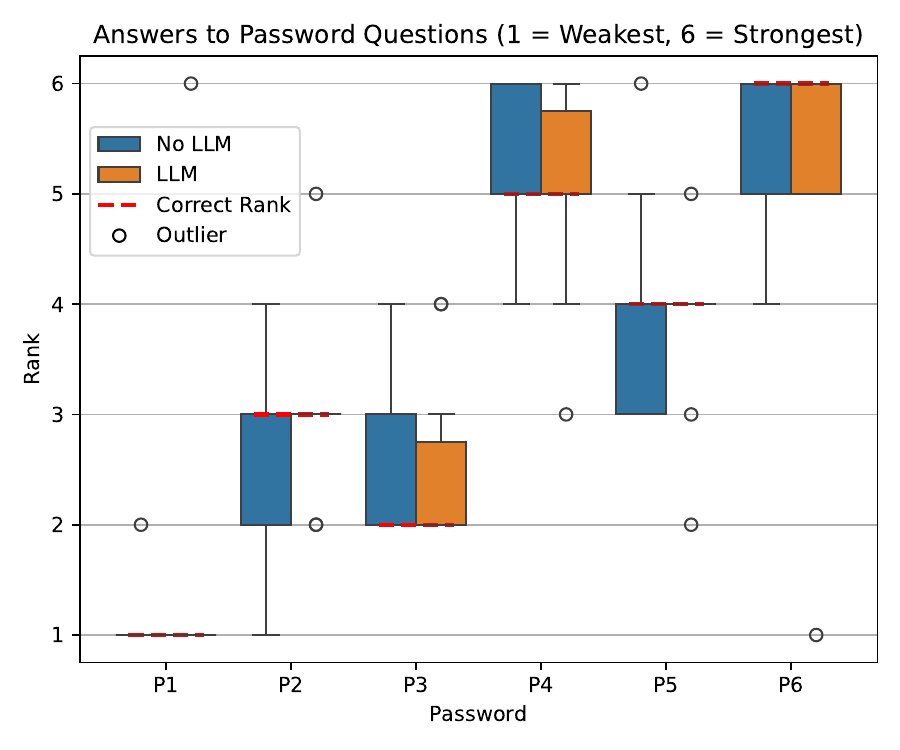}
    \caption{Boxplot of password strength rankings: \textmd{\footnotesize P1 = \texttt{123456}, P2 = \texttt{P@ssw0rd}, P3 = \texttt{qwerty123}, P4 = \texttt{s!XjGm@9t\#R2}, P5 = \texttt{MyDogCharlie2019}, P6 = \texttt{A3x\&zP!tQ@1K}; Red dashed lines indicate correct rankings. White circles with black borders denote outliers; 1 = weakest password, 6 = strongest password.}}
    \label{fig:password_strength_p1_to_p6}
\end{figure}

\noindent \textbf{Password Strength (Easy Difficulty) Questions} - Participants were also asked to rank six passwords from weakest to strongest (see Figure \ref{fig:password_strength_p1_to_p6}). This task further demonstrated the benefit of LLM support. Participants with an access to one of the LLMs produced rankings that aligned more closely with the  correct order, particularly at both extremes of the scale. The strongest password\footnote{\texttt{A3x\&zP!tQ@1K} is considered as a stronger password due to it having a more randomized character distribution, while \texttt{s!XjGm@9t\#R2} has more alternating letter cases that resemble common structured patterns.} (``P6'': \texttt{A3x\&zP!tQ@1K}) was frequently ranked correctly as the most secure by the LLM group, while the non-LLM group showed a wider spread of responses, indicating less certainty or awareness of password strength best practices. Similarly, both groups generally identified ``P1'' (\texttt{123456}) as the weakest password, but the LLM group demonstrated slightly higher consistency. The accuracy gap was more noticeable for mid-level passwords such as ``P3'' (\texttt{qwerty123}) and ``P5'' (\texttt{MyDogCharlie2019}) wherein common misconceptions about strength tend to arise. In general, the LLM group's rankings exhibited reduced variability, fewer outliers as well as closer adherence to expert expectations.

\begin{figure}[H]
    \centering
    \includegraphics[width=\columnwidth]{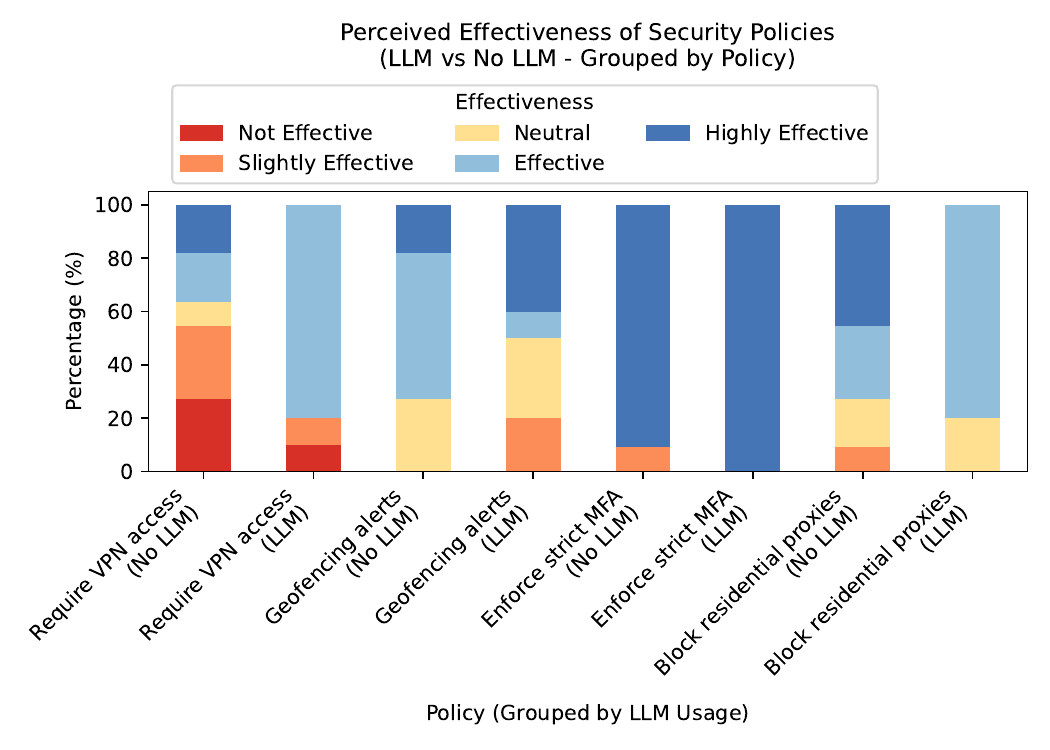}
    \caption{Perceived effectiveness of security policies for preventing unauthorized access.}
    \label{fig:policy-effectiveness-llm}
\end{figure}

\begin{figure}[H]
    \centering
    \includegraphics[width=\linewidth]{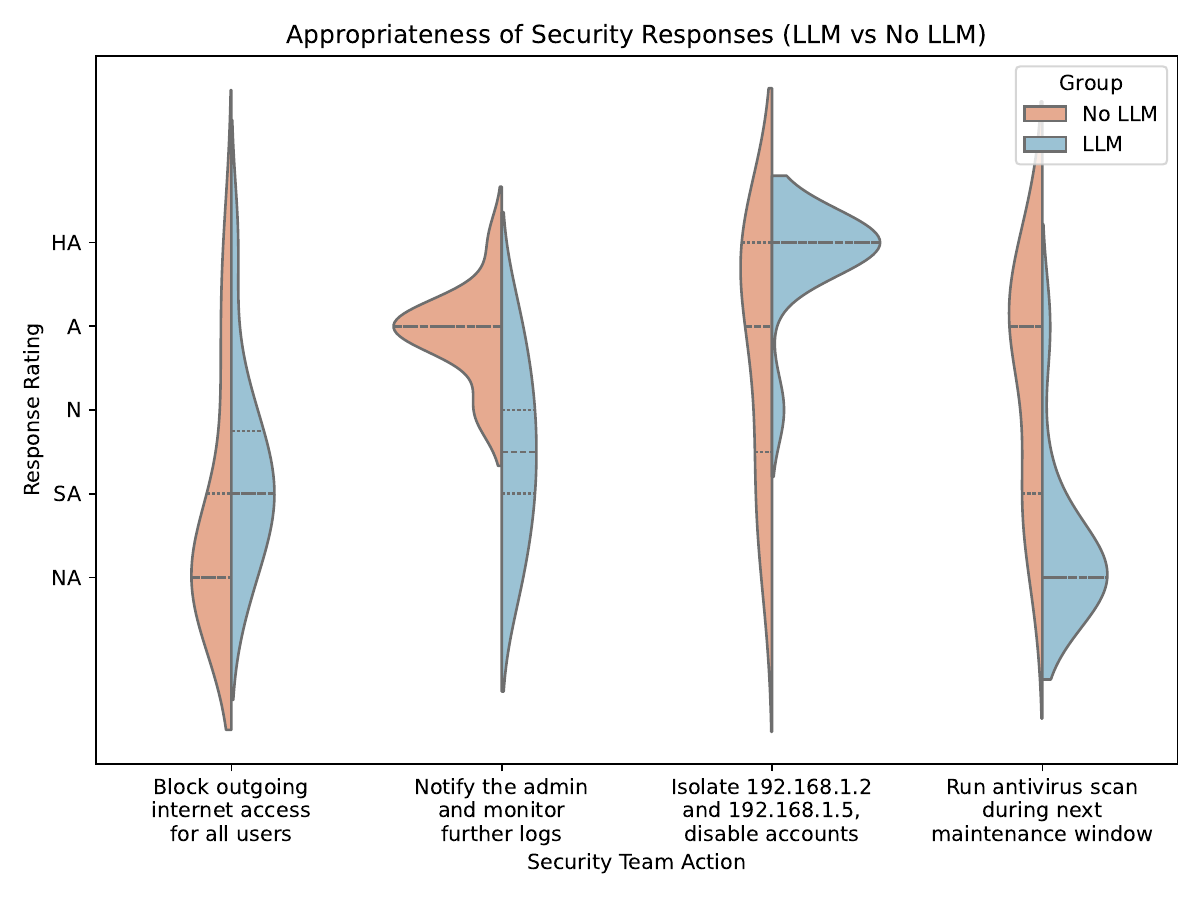}
    \caption{Perceived appropriateness of various security actions based on log evidence.}
    \label{fig:log_response_violin}
\end{figure}

\noindent \textbf{Policy Effectiveness (Medium Difficulty) Question} -  Participants also rated the effectiveness of four common cybersecurity policies. Once again, LLM-supported group was associated with more consistent evaluations (see Figure \ref{fig:policy-effectiveness-llm}). Across all four policies --- requiring VPN access, geofencing alerts, enforcing strict multi-factor authentication (MFA), and blocking residential proxies --- the LLM group reported higher proportions of ``Effective'' and ``Highly Effective'' ratings compared to their non-LLM counterparts. Notably, strict MFA and blocking residential proxies received particularly strong endorsements from the LLM group, with the majority rating these as ``Highly Effective''. In contrast, the non-LLM group exhibited greater variability, with more responses falling into the ``Neutral'' or ``Slightly Effective'' categories. This pattern suggests that participants with LLM assistance may have a clearer understanding of the role and impact of such policies or may have received contextual explanations that shaped their judgments.

\noindent \textbf{Appropriateness of Security Responses (Medium Difficulty) Question} - Participants were also asked to rate the appropriateness of four possible responses to a hypothetical security incident. Figure \ref{fig:log_response_violin} indicates that the presence of LLM support appeared to influence judgment patterns. The violin plot reveals that LLM-assisted participants were more consistent and more likely to rate appropriate and proportionate responses—such as isolating affected machines and disabling compromised accounts --- as ``Highly Appropriate''. In contrast, the non-LLM group exhibited more dispersed ratings, particularly for more aggressive or delayed responses such as blocking all outgoing internet access or waiting until the next maintenance window to run an antivirus scan. Notably, LLM users were more critical of extreme or delayed actions, favoring more targeted and timely interventions.

\begin{figure}[H]
    \centering
    \includegraphics[width=\linewidth]{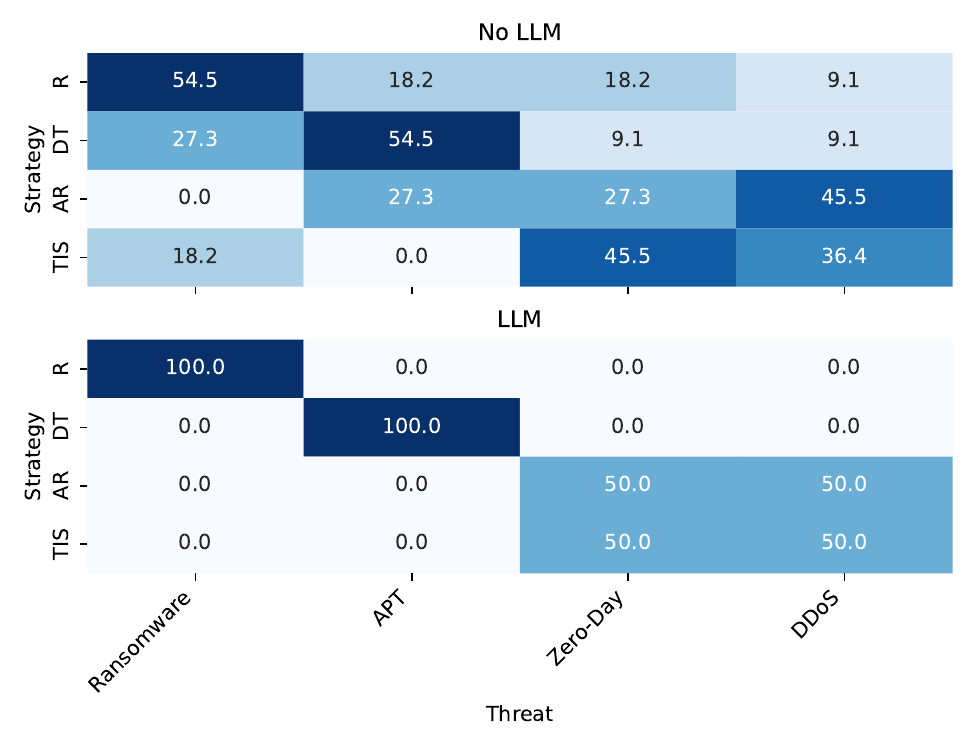}
    \caption{Comparison of perceived threat mitigation effectiveness by resilience strategy: \textmd{\footnotesize Rows represent strategies (R = Redundancy, DT = Deception Technology, AR = Automated Response Systems, TIS = Threat Intelligence Sharing); Values show percentage of participants who matched each strategy to a given threat.}}
    \label{fig:strategy-threat-matrix}
\end{figure}

\noindent \textbf{Threat Mitigation Effectiveness (Hard Difficulty) Question} - To further explore participants' cybersecurity reasoning, they were asked to match four defense strategies to four major threat types. The heatmap (see Figure \ref{fig:strategy-threat-matrix}) reveals marked differences in how the two groups allocated strategies such as redundancy, deception technology, automated response systems and threat intelligence sharing. The LLM-supported group demonstrated more structured and threat-aligned pairings. For instance, they more frequently matched ``Automated Response Systems'' with ``DDoS'' attacks and ``Deception Technology'' with ``APT'' threats, which are both widely accepted as strong strategic fits. In contrast, the non-LLM group exhibited more dispersed choices, with less consistent alignment across threats. However, the LLM group did not outperform across all pairings. Specifically, their allocation of ``Redundancy'' was less precise. While this strategy is typically well-suited for mitigating the impact of DDoS attacks or ransomware (by maintaining service availability), LLM participants often misapplied it to threats such as APTs and Zero-Day exploits (cases where redundancy offers limited protection).

\begin{figure}[H]
    \centering
    \includegraphics[width=\linewidth]{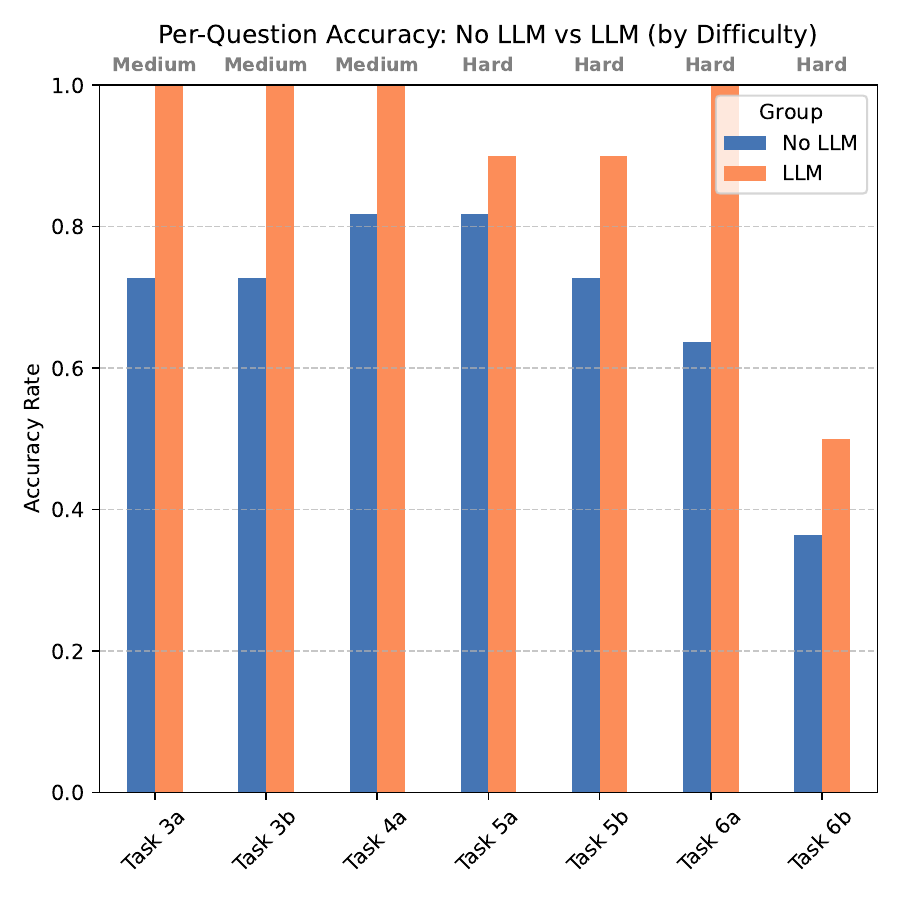}
    \caption{Single-choice Question Accuracy (with Difficulty Labels)}
    \label{fig:single_choice}
\end{figure}

\noindent \textbf{Single-Choice Questions} - Figure \ref{fig:single_choice} shows that overall accuracy varied by task difficulty and group. For the medium-difficulty items (Tasks 3a, 3b and 4a), both groups performed strongly, wherein the LLM group consistently selected the correct answers, thereby demonstrating clear benefit from targeted support. However, on the hard tasks (Tasks 5a, 5b, 6a and 6b), while both groups showed high accuracy on Tasks 5a and 5b, a divergence emerged on Task 6b. Although the LLM-supported group correctly answered Task 6a (identifying “Redundancy”), their performance on Task 6b, which required the selection of ``Threat Intelligence Sharing'', was notably inconsistent. Several LLM responses mistakenly opted for alternatives such as ``Automated Response Systems'' or ``Deception Technology''. This suggests that the guidance provided by the language model may have been less effective in clarifying nuanced distinctions among defense strategies for this particular task.

\subsection{Qualitative Analysis}

To further evaluate differences in cybersecurity reasoning, we conducted a thematic analysis of responses to open-ended questions from two participant groups. This analysis focused on three core tasks: phishing prevention, phishing containment, and password management. Responses were systematically coded using an inductive approach, wherein recurring ideas and concepts were grouped into themes without pre-defined categories. Two researchers independently reviewed the data (97\% intercoder reliability score) to identify patterns and then consolidated their codes through iterative discussions to ensure consistency and reduce bias. 

\begin{figure}[H]
    \centering
    \includegraphics[width=\linewidth]{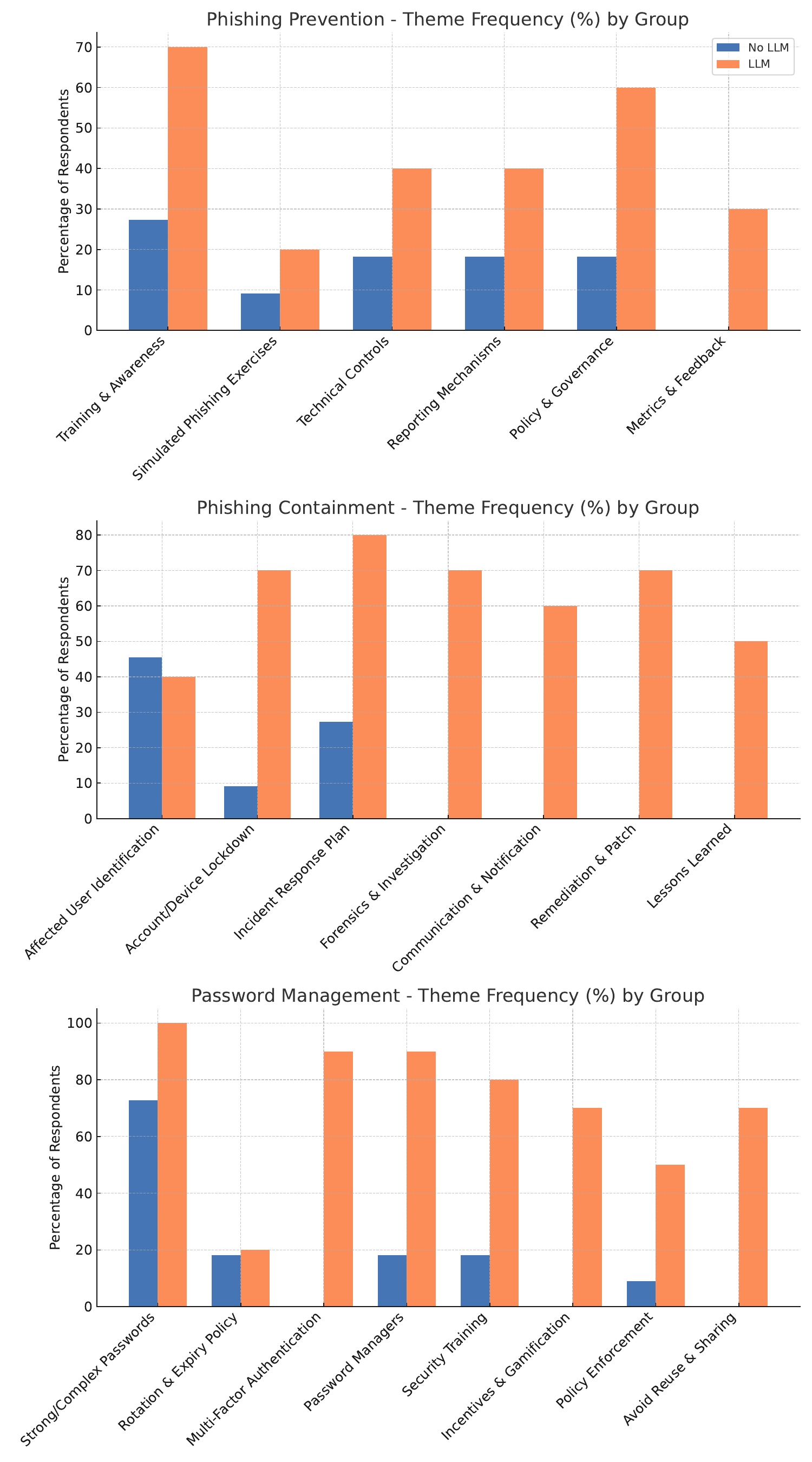}
    \caption{Comparison of theme frequency percentages for phishing prevention, containment and password management strategies.}
    \label{fig:theme_comparison}
\end{figure}

Figure \ref{fig:theme_comparison} demonstrated that LLM-supported group proposed broader and more structured strategies than their counterpart. For phishing prevention, 70\% of LLM responses emphasized training and awareness, compared to just 27\% among non-LLM users. LLM-generated responses also more frequently included technical controls, reporting mechanisms and policy governance, wherein feedback loops such as tracking the effectiveness of training were mentioned. In contrast, non-LLM participants focused more narrowly on conventional training and basic filtering. Regarding phishing containment, 70–80\% of LLM responses referenced specific containment actions, forensic procedures and remediation. Only the LLM group mentioned post-incident review and internal communication, which can be considered as formal cybersecurity protocols. Non-LLM responses tended to center on identifying affected users but rarely extended into containment or strategic follow-up. For password management, while both groups recognized the need for strong passwords, LLM users were far more likely to suggest multi-factor authentication (90\%), password managers (90\%) and training or incentive mechanisms to promote secure behavior. In contrast, the non-LLM group primarily focused on password complexity, often in isolation from broader behavioral or policy structures. 

Beyond the aforementioned themes, several non-LLM responses revealed valid but overlooked areas. Some advocated for visibility-based discipline, such as publicly showcasing the victim’s mistake: ``...showcase the victim’s actions in front of everybody to get maximum impact...'', which is a tactic absent from LLM outputs potentially due to ethical reasons. Others suggested biometric authentication (e.g., ``...login to the PC using Windows Hello with fingerprint or facescanner...''), highlighting a practical device-level security measure not addressed in the LLM’s password-centric themes. Finally, several participants mentioned specific technical integration practices such as ``...position email server in DMZ, reject non-SSL traffic...'', offering domain-specific details beyond the general ``technical controls'' commonly cited by LLMs.

\subsection{Impact of Personal Resilience}

To better understand the relationship between individual cybersecurity resilience and task performance, we conducted a correlation analysis using BRS values and participants’ task accuracy. We report both Pearson (linear) and Spearman (rank-order) correlation coefficients to capture different relationship structures and assess the robustness of associations across difficulty levels and LLM usage. This is presented in Table \ref{tab:brs_llm_colored}.

The analysis revealed several important patterns in the relationship between participants’ BRS scores and their performance on cybersecurity tasks. When aggregating all participants, there was a statistically significant positive correlation between BRS and task accuracy for both Pearson’s $r = 0.699$, $p = 0.001$ and Spearman’s $\rho = 0.747$, $p < 0.001$, suggesting that greater resilience was associated with higher overall performance.

\begin{table}[H]
\small
\centering
\resizebox{\columnwidth}{!}{%
\begin{tabular}{|l|l|c|c|c|c|}
\hline
\textbf{Task Group} & \textbf{LLM Use} & \textbf{Pearson $r$} & \textbf{$p$-value} & \textbf{Spearman $\rho$} & \textbf{$p$-value} \\
\hline
Overall  & Both & \cellcolor{palegreen}0.699 & \cellcolor{palegreen}0.001 & \cellcolor{palegreen}0.747 & \cellcolor{palegreen}0.000 \\
Easy     & Both & 0.069 & 0.767 & 0.260 & 0.256 \\
Medium   & Both & \cellcolor{palegreen}0.662 & \cellcolor{palegreen}0.001 & \cellcolor{palegreen}0.705 & \cellcolor{palegreen}0.000 \\
Hard     & Both & 0.401 & 0.072 & 0.413 & 0.063 \\
\hline
\multirow{2}{*}{Overall} & No LLM & 0.491 & 0.125 & 0.564 & 0.071 \\
                         & LLM    & \cellcolor{palegreen}0.864 & \cellcolor{palegreen}0.001 & \cellcolor{palegreen}0.924 & \cellcolor{palegreen}0.000 \\
\hline
\multirow{2}{*}{Easy}    & No LLM & -0.409 & 0.211 & -0.320 & 0.338 \\
                         & LLM    & \cellcolor{palegreen}0.717 & \cellcolor{palegreen}0.020 & \cellcolor{palegreen}0.846 & \cellcolor{palegreen}0.002 \\
\hline
\multirow{2}{*}{Medium}  & No LLM & 0.511 & 0.108 & \cellcolor{palegreen}0.619 & \cellcolor{palegreen}0.042 \\
                         & LLM    & \cellcolor{palegreen}0.839 & \cellcolor{palegreen}0.002 & \cellcolor{palegreen}0.801 & \cellcolor{palegreen}0.005 \\
\hline
\multirow{2}{*}{Hard}    & No LLM & \cellcolor{palegreen}0.578 & 0.063 & \cellcolor{palegreen}0.699 & \cellcolor{palegreen}0.017 \\
                         & LLM    & 0.256 & 0.475 & 0.313 & 0.378 \\
\hline
\end{tabular}
}
\caption{Correlations between BRS scores and task performance: \textmd{\footnotesize Significant values ($p < 0.05$) are highlighted in pale green.}}
\label{tab:brs_llm_colored}
\end{table}

When splitting the tasks by difficulty, a significant correlation emerged for medium-difficulty tasks, with both Pearson and Spearman values exceeding 0.66 and $p$-values below 0.001. This indicates that participants with higher resilience consistently performed better on moderately complex scenarios. No significant correlation was observed for easy tasks, possibly due to a ceiling effect and the hard task set showed moderate associations (not statistically significant).

Further analysis distinguished between unaided vs LLM-supported groups. Results show stark differences: the LLM group demonstrated very strong and statistically significant correlations between BRS and task performance across all task types (e.g., Spearman’s $\rho = 0.924$, $p < 0.001$ overall). In contrast, the No LLM group displayed moderate correlations that did not reach significance, with the exception of hard tasks, where Spearman’s $\rho = 0.699$, $p = 0.017$ suggested a noteworthy link between resilience and success in highly complex situations.

Participants did not always utilize the available LLM support, even when it was permitted and potentially beneficial. While LLM usage was nearly universal in tasks such as password strength evaluation and mitigation strategy mapping (e.g., Task 5a and 6a), a subset of participants chose to respond without LLM assistance in more cognitively demanding tasks like threat attribution (Task 4a), incident containment (Task 3b), and evaluation of response appropriateness based on log data. This selective non-use may reflect factors such as overconfidence in personal judgment, time pressure, or uncertainty regarding the scope of LLM use.

Across these tasks, LLM usage was consistently associated with higher answer accuracy. For instance, participants who used the LLM in Task 3b (choosing an appropriate containment response) and Task 4a (explaining a sequence of suspicious events) were more likely to select specific, technically sound actions such as revoking credentials or identifying privilege escalation. In contrast, those who answered without LLM support frequently selected vague or less effective responses, such as deferring action or monitoring logs without intervention. These findings suggest that LLMs functioned as effective reasoning aids, particularly for tasks requiring multi-step inference, security expertise, and alignment with best practices.

\section{Discussion}
\label{sec:discussion}

In the following, we present the key findings, discuss their implications, outline future research directions and note the limitations of our study.

\subsection{Key Findings and Implications}

Our study reveals that LLMs have a clear influence on cybersecurity decision-making. However, this influence is by no means uniform. While LLMs consistently improved performance across easy and medium tasks, particularly in tasks such as phishing detection and password evaluation, they also introduced complex behavioral effects that raise questions on the assumptions of universal benefit.

In regards to the interaction between BRS and task accuracy (especially in the presence of LLM support), the correlation patterns between BRS and task accuracy reveal that LLMs may act as performance amplifiers for individuals with high cognitive resilience. In the LLM-supported group, resilient participants achieved higher performance across all task difficulties. This pattern was statistically significant and robust, suggesting that those with greater mental adaptability, persistence, and problem-solving capacity are better equipped to harness LLM capabilities.

In contrast, the participants of the unaided group showed only moderate or task-specific correlations between resilience and performance. It is worth noting that for the most difficult tasks, high-resilience individuals without LLMs sometimes outperformed their LLM-supported peers. This points to a potential cognitive tradeoff in complex scenarios where ambiguity is high and judgment is critical, as LLMs may introduce misleading suggestions, ambiguity, or irrelevant information added negatives that distract from core reasoning. These findings align with broader concerns about automation bias and cognitive offloading, wherein users may place excessive trust in AI-generated output at the expense of their own evaluative reasoning.

Even in easy tasks, we observed that the LLM improved the resilience-performance link, suggesting that model assistance reinforces existing strengths but does not compensate for weaknesses. Less resilient individuals derived fewer benefits from LLM usage, and in some cases, their performance stagnated or even declined. This suggests a troubling implication that without proper design, LLMs may inadvertently widen performance disparities by disproportionately benefiting users who are already more adept at critical thinking or stress regulation. LLMs may highlight inequalities in decision accuracy and reliability, particularly in high-stakes domains such as cybersecurity.

Beyond performance metrics, our qualitative analysis revealed that LLMs also shape the structure and depth of reasoning. LLM-supported participants provided broader and more systematic responses, incorporating layered strategies and best practices. However, these benefits were less apparent in edge cases or important situations requiring domain-specific expertise. In some instances, LLM-supported users offered technically plausible but contextually inappropriate responses, suggesting that model fluency can mask a lack of understanding. This leads users to accept confident-sounding yet incorrect outputs. Furthermore, this raises important risks for over-reliance, especially when users are not sufficiently equipped to evaluate or cross-check model suggestions.

These patterns support an emerging hypothesis that LLMs are most beneficial when paired with cognitively resilient users who can critically evaluate and adapt model outputs. This raise a critical question for the design of LLM-integrated systems: \emph{How can we ensure that users with varying levels of resilience benefit equally?} One implication is that LLMs should be embedded within interactive, adaptive interfaces that adjust explanations, feedback, or decision scaffolds based on user profile—mirroring the concept of “human-aligned AI” in high-stakes workflows.

While LLMs show clear promise as decision aids, they also carry inherent risks that range from inducing overconfidence and narrowing cognitive diversity to masking errors (e.g., hallucination) through ``superficially fluent output.'' These limitations must be addressed not only through better training and transparency but through systems-level design that foregrounds human judgment, diversity and autonomy.

Our findings suggest several design implications for human-in-the-loop AI systems. First, interfaces should adapt to users' cognitive profiles (especially resilience levels) by adapting the depth of LLM support. For example, high-resilience users may benefit from open-ended suggestions, while lower-resilience users might require clearer guidance, confidence indicators or error warnings to avoid blind trust. Second, to counter the convergence effects observed in our thematic analysis, systems can incorporate scaffolding prompts that encourage users to consider multiple perspectives or challenge default assumptions (e.g., ``What would a legal, ethical, or adversarial viewpoint say?''). Finally, reflective feedback loops—such as brief prompts asking users to explain or reassess their decisions can promote critical engagement and reduce passive over-reliance. 

Rather than treating the LLM as a static answer engine, future systems should frame it as an adaptive collaborator, which can promote judgment diversity and preserve human agency in security-critical workflows.

\subsection{Future Work}

As part of our future work, we plan to expand the scale and diversity of participant populations. Our sample consisted of graduate students with relatively high baseline (cybersecurity) technical competence. To generalize findings we will include also cybersecurity professionals, IT generalists and non-experts that are part of security-critical tasks within their organizations. Investigating whether there is an improvement when participants utilize LLMs across different skill levels and job roles will help explain potential observed interaction boundaries.

Furthermore, it would be beneficial to explore adaptive LLM interfaces that dynamically target their support based on users’ cognitive and behavioral profiles. Instead of offering static outputs, LLM tools could adjust their level of detail, explanation and autonomy in real time based on the participant. For example, systems could detect signs of over-reliance or under-confidence (e.g., through user hesitancy or interaction patterns) and respond with different types of answers.

Another potential direction involves investigating whether LLM use can shape resilience itself, rather than merely interacting with it. Longitudinal research could explore whether repeated exposure to LLM-assisted tasks strengthens or weakens individuals' ability to not repeat certain errors. Moreover, future studies should include a pure LLM group. This condition would serve as a valuable baseline for evaluating human and AI differences. It would also allow us to systematically compare different LLMs (e.g., GPT-4, Claude and various open-source models) under consistent task conditions, which would provide insights on model-specific strengths and weaknesses.

Lastly, we also see benefits in extending our study beyond accuracy to include outcomes such as trust calibration, perceived control, ethical awareness and collaborative behaviors in multi-agent decision contexts. This is because LLMs are increasingly embedded in teams and organizational workflows and understanding their impact on group-level dynamics (especially under stress or attack scenarios) is crucial.

\subsection{Limitations}

Our sample size was relatively small (N = 21) and recruited through purposive sampling. Although the participants were well-suited for an exploratory study (master's students with training in information security) the limited sample restricts broader claims about LLM impact across industries, age groups, experience levels and cultures. The high variance in baseline resilience scores between focus groups also means that some observed effects may reflect group composition rather than general trends. However, small-sample studies are common in cybersecurity research due to the challenges involved in recruiting larger numbers of suitable participants (\cite{braun2024understanding}).

Furthermore, the focus group methodology is subject to social dynamics that may create some bias in results. Participants’ reasoning processes and decisions may have been influenced by the group setting or the presence of facilitators. However, we attempted to mitigate this through structured individual tasks as well as anonymized response logging. In regards to the BRS, it was designed to capture general psychological resilience but does not address domain-specific facets such as security expertise, decision fatigue under time pressure or emotional resilience to stressors common in cybersecurity roles. In future, it would be beneficial to consider multi-dimensional resilience measures that incorporate situational, emotional and cognitive components.

Moreover, the classification of task difficulty may not fully capture individual perceptions of complexity or ambiguity. That means that tasks that are considered easy for one participant may be difficult for another participants. This introduces potential noise into our analysis of difficulty-resilience interactions and highlights the need for more personalized task modeling in future experiments. We mitigated this limitation by incorporating tasks similar to those found in standard cybersecurity certification exams (e.g. CompTIA Security+). This ensured that the activities were both realistic and relevant to professional practice.

Finally, while we observed significant correlations between resilience and task performance in the LLM group, it is not possible to infer causality from such data. There exist the chance that other unmeasured variables (prior familiarity with LLMs, cognitive style or interface usability) could have potentially influenced the outcomes. Possible directions how this aspect could be addressed involves application of resilience priming, feedback conditions or LLM usage constraints.

\section{Conclusion}
\label{sec:conclusion}

This study explored how LLM support influences human decision-making in cybersecurity tasks of varying complexity. While LLMs improve task accuracy, especially in routine scenarios, their benefits are not evenly distributed. Cognitive resilience plays a pivotal role in determining whether users can critically engage with or become over-reliant on AI suggestions. Our findings suggest that to realize the full potential of human–LLM interaction, future systems must adapt to user capabilities, preserve decision diversity as well as mitigate risks of automation bias. Ensuring that LLMs augment rather than diminish human judgment will be central to their responsible integration in security-critical workflows.

\paragraph*{Acknowledgments}
We thank the Hilti Corporation for funding, and the students we interviewed for their contributions, availability, and valuable feedback.

\renewcommand*{\bibfont}{\small}
\printbibliography

@article{zhang2025llms,
  title={When llms meet cybersecurity: A systematic literature review},
  author={Zhang, Jie and Bu, Haoyu and Wen, Hui and Liu, Yongji and Fei, Haiqiang and Xi, Rongrong and Li, Lun and Yang, Yun and Zhu, Hongsong and Meng, Dan},
  journal={Cybersecurity},
  volume={8},
  number={1},
  pages={1--41},
  year={2025},
  publisher={SpringerOpen}
}

@article{ding2024large,
  title={Large Language Models for Cyber Resilience: A Comprehensive Review, Challenges, and Future Perspectives},
  author={Ding, Weiping and Abdel-Basset, Mohamed and Ali, Ahmed M and Moustafa, Nour},
  journal={Applied Soft Computing},
  pages={112663},
  year={2024},
  publisher={Elsevier}
}

@article{khadka2025human,
  title={Human factors in cybersecurity: an interdisciplinary review and framework proposal},
  author={Khadka, Kalam and Ullah, Abu Barkat},
  journal={International Journal of Information Security},
  volume={24},
  number={3},
  pages={1--13},
  year={2025},
  publisher={Springer}
}

@article{cohen2025human,
  title={Human--AI Enhancement of Cyber Threat Intelligence},
  author={Cohen, Daniel and Te’eni, Dov and Yahav, Inbal and Zagalsky, Alexey and Schwartz, David and Silverman, Gahl and Mann, Yossi and Elalouf, Amir and Makowski, Jeremy},
  journal={International Journal of Information Security},
  volume={24},
  number={2},
  pages={99},
  year={2025},
  publisher={Springer}
}

@article{bedoya2024enhancing,
  title={Enhancing DevSecOps practice with Large Language Models and Security Chaos Engineering},
  author={Bedoya, Martin and Palacios, Sara and D{\'\i}az-L{\'o}pez, Daniel and Laverde, Estefania and Nespoli, Pantaleone},
  journal={International Journal of Information Security},
  pages={1--24},
  year={2024},
  publisher={Springer}
}

@article{sarker2024explainable,
  title={Explainable AI for cybersecurity automation, intelligence and trustworthiness in digital twin: Methods, taxonomy, challenges and prospects},
  author={Sarker, Iqbal H and Janicke, Helge and Mohsin, Ahmad and Gill, Asif and Maglaras, Leandros},
  journal={ICT Express},
  year={2024},
  publisher={Elsevier}
}

@inproceedings{nguyen2024utilizing,
  title={Utilizing large language models with human feedback integration for generating dedicated warning for phishing emails},
  author={Nguyen, Quan Hong and Wu, Tingmin and Nguyen, Van and Yuan, Xingliang and Xue, Jason and Rudolph, Carsten},
  booktitle={Proceedings of the 2nd ACM Workshop on Secure and Trustworthy Deep Learning Systems},
  pages={35--46},
  year={2024}
}

@article{chen2024survey,
  title={A survey of large language models for cyber threat detection},
  author={Chen, Yiren and Cui, Mengjiao and Wang, Ding and Cao, Yiyang and Yang, Peian and Jiang, Bo and Lu, Zhigang and Liu, Baoxu},
  journal={Computers \& Security},
  pages={104016},
  year={2024},
  publisher={Elsevier}
}

@article{fuegener2021borgs,
	title        = {{Will Humans-in-the-Loop Become Borgs? Merits and Pitfalls of Working with AI}},
	author       = {F{\"u}gener, Andreas and Grahl, J{\"o}rn and Gupta, Alok and Ketter, Wolfgang},
	year         = 2021,
	journal      = {MIS Quarterly},
	publisher    = {MISRC},
	volume       = 45,
	number       = 3,
	pages        = {1527--1556}
}

@article{shen2025irrationality,
	title        = {{Irrationality-Aware Human Machine Collaboration: Mitigating Alterfactual Irrationality in Copy Trading}},
	author       = {Shen, Zhe and Jiang, Wei and Zheng, Zhiqiang},
	year         = 2025,
	journal      = {Information Systems Research},
	publisher    = {INFORMS}
}

@article{lu2024information,
	title        = {{Information, Humans, and Machines}},
	author       = {Lu, Tianyang and Zhang, Yuqian},
	year         = 2025,
	journal      = {Information Systems Research},
	publisher    = {INFORMS},
	volume       = 36,
	number       = 1,
	pages        = {394--418}
}

@incollection{gallagher2024phishing,
	title        = {{Phishing and social engineering in the age of llms}},
	author       = {Gallagher, Sean and Gelman, Ben and Taoufiq, Salma and V{\"o}r{\"o}s, Tam{\'a}s and Lee, Younghoo and Kyadige, Adarsh and Bergeron, Sean},
	year         = 2024,
	booktitle    = {Large Language Models in Cybersecurity: Threats, Exposure and Mitigation},
	publisher    = {Springer Nature Switzerland Cham},
	pages        = {81--86}
}

@article{jacobs2021antidepressants,
	title        = {{How Machine-Learning Recommendations Influence Clinician Treatment Selections: The Example of Antidepressant Selection}},
	author       = {Jacobs, Michelle and Pradier, Marie-France and McCoy, Thomas H and Perlis, Roy H and Doshi-Velez, Finale and Gajos, Krzysztof Z},
	year         = 2021,
	journal      = {Translational Psychiatry},
	volume       = 11,
	number       = 1,
	pages        = {1--9}
}

@article{jussupow2021diagnosis,
	title        = {{Augmenting Medical Diagnosis Decisions? An Investigation into Physicians’ Decision-Making Process with Artificial Intelligence}},
	author       = {Jussupow, Egor and Spohrer, Kai and Heinzl, Armin and Gawlitza, Jan},
	year         = 2021,
	journal      = {Information Systems Research},
	volume       = 32,
	number       = 3,
	pages        = {713--735}
}

@book{decremer2021kasparov,
	title        = {{The Future of Work: How Artificial Intelligence Can Complement Human Decision-Making}},
	author       = {De Cremer, David and Kasparov, Garry},
	year         = 2021,
	publisher    = {Harvard Business Review Press}
}

@book{page2007diversity,
	title        = {{The Difference: How the Power of Diversity Creates Better Groups, Firms, Schools, and Societies}},
	author       = {Page, Scott E},
	year         = 2007,
	publisher    = {Princeton University Press}
}

@article{hong2004groups,
	title        = {{Groups of diverse problem solvers can outperform groups of high-ability problem solvers}},
	author       = {Hong, Lu and Page, Scott E},
	year         = 2004,
	journal      = {Proceedings of the National Academy of Sciences},
	volume       = 101,
	number       = 46,
	pages        = {16385--16389}
}

@article{smith2008brief,
	title        = {The brief resilience scale: assessing the ability to bounce back},
	author       = {Smith, Bruce W and Dalen, Jeanne and Wiggins, Kathryn and Tooley, Erin and Christopher, Paulette and Bernard, Jennifer},
	year         = 2008,
	journal      = {International journal of behavioral medicine},
	publisher    = {Springer},
	volume       = 15,
	pages        = {194-200}
}

@article{morgan1996focusgroups,
	title        = {{Focus groups}},
	author       = {{Morgan, D.L.}},
	year         = 1996,
	journal      = {{Annual review of sociology}},
	volume       = 22,
	number       = 1,
	pages        = {129-152}
}

@inproceedings{braun2024understanding,
  title={Understanding the Process of Data Labeling in Cybersecurity},
  author={Braun, Tobias and Pekaric, Irdin and Apruzzese, Giovanni},
  booktitle={Proceedings of the 39th ACM/SIGAPP Symposium on Applied Computing},
  pages={1596--1605},
  year={2024}
}

@inproceedings{pekaric2025mobile,
  title={How Do Mobile Applications Enhance Security? An Exploratory Analysis of Use Cases and Provided Information},
  author={Pekaric, Irdin and Sauerwein, Clemens and Laichner, Simon and Breu, Ruth},
  booktitle={Proceedings of the 2025 ACM Southeast Conference},
  pages={114--123},
  year={2025}
}

@article{schroer2025dark,
  title={The Dark Side of the Web: Towards Understanding Various Data Sources in Cyber Threat Intelligence},
  author={Schr{\"o}er, Saskia Laura and Canevascini, No{\'e} and Pekaric, Irdin and Widmer, Philine and Laskov, Pavel},
  journal={arXiv preprint arXiv:2504.14235},
  year={2025}
}

@inproceedings{witte2022towards,
  title={Towards model co-evolution across self-adaptation steps for combined safety and security analysis},
  author={Witte, Thomas and Groner, Raffaela and Raschke, Alexander and Tichy, Matthias and Pekaric, Irdin and Felderer, Michael},
  booktitle={Proceedings of the 17th Symposium on Software Engineering for Adaptive and Self-Managing Systems},
  pages={106--112},
  year={2022}
}

@article{pekaric2021vulnerlizer,
  title={VULNERLIZER: Cross-analysis Between Vulnerabilities and Software Libraries},
  author={Pekaric, Irdin and Felderer, Michael and Steinm{\"u}ller, Philipp},
  year={2021}
}

@inproceedings{groner2023model,
  title={Model-based generation of attack-fault trees},
  author={Groner, Raffaela and Witte, Thomas and Raschke, Alexander and Hirn, Sophie and Pekaric, Irdin and Frick, Markus and Tichy, Matthias and Felderer, Michael},
  booktitle={International Conference on Computer Safety, Reliability, and Security},
  pages={107--120},
  year={2023},
  organization={Springer}
}

\end{document}